\documentclass[sigconf]{acmart}
\usepackage{booktabs}
\usepackage{listings}
\usepackage{hyperref}
\usepackage[shortlabels]{enumitem}
\usepackage{listings}

\definecolor{codegreen}{rgb}{0,0.6,0.2}
\definecolor{codegray}{rgb}{0.5,0.5,0.5}
\definecolor{codeblue}{rgb}{0,0,0.8}
\definecolor{backcolour}{rgb}{0.95,0.95,0.92}

\lstdefinestyle{mystyle1}{
    backgroundcolor=\color{backcolour},
    commentstyle=\color{codegreen},
    keywordstyle=\color{blue},
    numberstyle=\tiny\color{codegray},
    stringstyle=\color{red},
    basicstyle=\ttfamily\small,
    breakatwhitespace=false,
    breaklines=true,
    captionpos=b,
    keepspaces=true,
    numbers=left,
    numbersep=5pt,
    showspaces=false,
    showstringspaces=false,
    showtabs=false,
    tabsize=2
}

\lstdefinestyle{mystyle2}{
    backgroundcolor=\color{backcolour},
    commentstyle=\color{codegreen},
    keywordstyle=\color{black},
    numberstyle=\tiny\color{codegray},
    stringstyle=\color{red},
    basicstyle=\ttfamily\small,
    breakatwhitespace=false,
    breaklines=true,
    captionpos=b,
    keepspaces=true,
    numbers=left,
    numbersep=5pt,
    showspaces=false,
    showstringspaces=false,
    showtabs=false,
    tabsize=2
}

\AtBeginDocument{%
  \providecommand\BibTeX{{%
    \normalfont B\kern-0.5em{\scshape i\kern-0.25em b}\kern-0.8em\TeX}}}

\setcopyright{rightsretained}
\copyrightyear{2020}
\acmYear{2020}
\acmDOI{10.5281/zenodo.3736363}
\acmISBN{}

\acmConference[ELS'20]{the 13th European Lisp Symposium}{April 27--28
  2020}{Z\"urich, Switzerland}

\begin{document}

\title{Bringing GNU Emacs to Native Code}

\author{Andrea Corallo}
\email{akrl@sdf.org}
\affiliation{}

\author{Luca Nassi}
\email{luknax@sdf.org}
\affiliation{}

\author{Nicola Manca}
\email{nicola.manca@spin.cnr.it}
\affiliation{
  \institution{CNR-SPIN}
  \streetaddress{Corso Perrone 24}
  \city{Genoa}
  \state{Italy}
  \postcode{16152}
}
\orcid{0000-0002-7768-2500}

\lstset{stringstyle=\ttfamily, basicstyle=\small, showstringspaces=false}

\begin{abstract}
  Emacs Lisp (Elisp) is the Lisp dialect used by the Emacs text editor
  family.  GNU Emacs can currently execute Elisp code either
  interpreted or byte-interpreted after it has been compiled to
  byte-code.  In this work we discuss the implementation of an
  optimizing compiler approach for Elisp targeting native code.  The
  native compiler employs the byte-compiler's internal representation
  as input and exploits \textit{libgccjit} to achieve code generation
  using the GNU Compiler Collection (GCC) infrastructure.  Generated
  executables are stored as binary files and can be loaded and
  unloaded dynamically. Most of the functionality of the compiler is
  written in Elisp itself, including several optimization passes,
  paired with a C back-end to interface with the GNU Emacs core and
  \textit{libgccjit}.  Though still a work in progress, our
  implementation is able to bootstrap a functional Emacs and compile
  all lexically scoped Elisp files, including the whole GNU Emacs Lisp
  Package Archive (ELPA)~\cite{ELS:ELPA}.  Native-compiled Elisp shows
  an increase of performance ranging from 2.3x up to 42x with respect
  to the equivalent byte-code, measured over a set of small
  benchmarks.
\end{abstract}



\maketitle

\section{Introduction}

GNU Emacs is known as the extensible, customizable, free/libre text
editor~\cite{ELS:EMACS_home}. This is not only one of the most iconic
text editors, GNU Emacs (from now on just ``Emacs'' for simplicity)
represents metaphorically the hearth of the GNU operating system.
Emacs can be described as a Lisp implementation (Emacs Lisp) and a
very broad set of Lisp programs written on top that, capable of a
surprising variety of tasks.  Emacs' design makes it one of the most
popular Lisp implementations to date.  Despite being widely employed,
Emacs has maintained a remarkably na\"{i}ve design for such a
long-standing project.  Although this makes it didactic, some
limitations prevent the current implementation of Emacs Lisp to be
appealing for broader use. In this context, performance issues
represent the main bottleneck, which can be broken down in three main
sub-problems:
\begin{itemize}
\item lack of true multi-threading support,
\item garbage collection speed,
\item code execution speed.
\end{itemize}
From now on we will focus on the last of these issues, which
constitutes the topic of this work.

The current implementation traditionally approaches the problem of
code execution speed in two ways:
\begin{itemize}
\item Implementing a large number of performance-sensitive primitive
  functions (also known as {\emph{subr}}) in C.
\item Compiling Lisp programs into a specific assembly representation
  suitable for targeting the Emacs VM called Lisp Assembly Program
  (LAP) and assembling it into byte-code.  This can be eventually
  executed by the
  byte-interpreter~\cite[Sec.1.2]{ELS:ELISP_bytecode},~\cite[Sec.~5.1]{ELS:Elisp_evo}.
\end{itemize}
As a result, Emacs developers had to implement a progressively
increasing amount of functions as C code primarily for performance
reasons.  As of Emacs~25, 22\% of the codebase was written in
C~\cite[Sec.~1.1]{ELS:ELISP_bytecode}, with consequences on
maintainability and extensibility~\cite{ELS:compilador}.  The last
significant performance increase dates back to around 1990, when an
optimizing byte-compiler including both source level and byte-code
optimizations was merged from Lucid
Emacs~\cite[Sec.~7.1]{ELS:Elisp_evo}. However, despite progressive
improvements, the main design of the byte-code machine stands
unmodified since then.  More recently, the problem of reaching better
performance has been approached using Just-In-Time (JIT) compilation
techniques, where three such implementations have been attempted or
proposed so far~\cite[Sec.~5.11]{ELS:Elisp_evo},
\cite{ELS:Tromey_blog}.  Possibly due to their simplistic approaches
none of them proved to introduce sufficient speed-up, in particular if
compared to the maintenance and dependency effort to be included in
the codebase.  In contrast, state-of-the-art high-performance Lisp
implementations rely on optimizing compilers targeting native code to
achieve higher performance~\cite{ELS:MacLachlan1992}.  In this
context, C-derived toolchains are already employed by a certain
number of Common Lisp implementations derived from KCL~\cite{ELS:KCL,
  ELS:GCL, ELS:Attardi1995, ELS:CLASP2015, ELS:CLASP2018}, where all
these, except CLASP, target C code generation.

In this work we present a different approach to tackle this problem,
based on the use of a novel intermediate representation (IR) to bridge
Elisp code with the GNU Compiler Collection~\cite{ELS:GCC_home}.  This
intermediate representation allows to effectively implement a number
of optimization passes and for Elisp byte-code to be translated to a
C-like semantic, compatible with the full pipeline of GCC optimization
passes.  This process relies on \textit{libgccjit} to plug into the
GCC infrastructure and achieve code generation without having to
target any intermediate programming language~\cite{ELS:libgccjit}.
The result of the compilation process for a compilation unit (CU) is a
file with \verb|.eln| extension (Emacs Lisp Native).  This is a new
file extension we have defined to hold the generated code and all the
necessary data to have it re-loadable over different Emacs runs.  This
last characteristic, in contrast with typical JIT-based approaches,
saves from having to recompile the same code at each run and allows
for more time expensive optimization passes.  Also, the classical Lisp
image dump feature is supported.

From a more general point of view, here we demonstrate how a Lisp
implementation can be hosted on top of \textit{libgccjit}.  Although
different libraries for code generation, such as
\textit{libjit}~\cite{ELS:LibJIT} or LLVM~\cite{ELS:LLVM}, have been
successfully employed by various Lisp implementations so
far~\cite{ELS:guile, ELS:CLASP2018}, we are not aware of any
leveraging \textit{libgccjit}.  Moreover, the proposed infrastructure
introduces better support for functional programming style in Emacs
Lisp with a pass performing tail recursion
elimination~\cite{ELS:wiki_tail_call} and the capability to be further
extended in order to perform full tail call optimization.

\begin{figure}
  \includegraphics[]{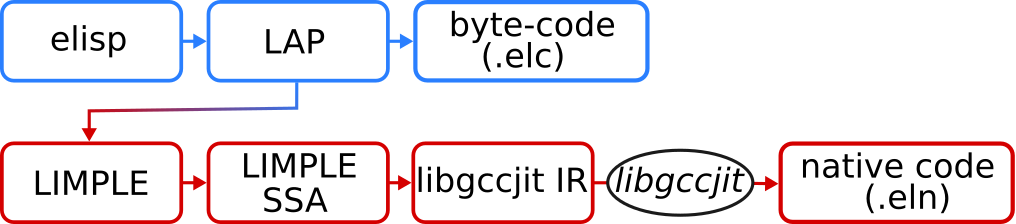}
  \caption{Program representation formats used by the byte-compiler
    (blue) and the native compiler (red) pipelines.}
  \label{fig1}
  \vspace{-0.5cm}
\end{figure}

\section{Implementation}

The proposed compiler pipeline can be divided in three main stages:
\begin{itemize}
\item Front-end: Lisp programs are compiled into LAP by the current
  byte-compiler.
\item Middle-end: LAP is converted into ``LIMPLE'', a new intermediate
  representation named after GCC GIMPLE~\cite{ELS:GIMPLE} which is the
  very core of the proposed compiler infrastructure.  LIMPLE is a
  sexp-based IR used as static single assignment (SSA)
  representation~\cite{ELS:SSA_book,ELS:wiki_SSA}.  Middle-end
  passes manipulate LIMPLE by performing a series of transformations
  on it.
\item Back-end: LIMPLE is converted into the \textit{libgccjit} IR to
  trigger the final compilation through the conventional GCC pipeline.
\end{itemize}
The sequence of program representation formats is presented in
Figure~\ref{fig1}.  The compiler takes care of type and value
propagation through the program control flow graph.  We point out
that, since Emacs Lisp received in 2012 lexical scope support, two
different sub-languages are currently
coexisting~\cite[Sec.~8.1]{ELS:Elisp_evo}.  The proposed compiler
focuses on generating code for the new lexically scoped dialect only,
since the dynamic one is considered obsolete and close to deprecation.

\subsection{LAP to \textit{libgccjit} IR}

Here, we briefly discuss the two endpoints of our compilation
pipeline: the Lisp Assembly Program and the \textit{libgccjit} IR, by
showing different representations of an illustrative and simple
code.

LAP representation is a list of instructions and labels, expressed in
terms of S-expressions. Instructions are assembled into byte-code,
where each one is associated with an operation and a manipulation of
the execution stack, both happening at runtime and defined by the
opcode corresponding to the instruction. When present, control flow
instructions can cause a change in the execution flow by executing
jumps to labels.  As an example, the Lisp expression
\lstinline[basicstyle=\ttfamily]{(if *bar* (+ *bar* 2) 'foo)}
is compiled into the following LAP representation:
\begin{lstlisting}[language=lisp, style=mystyle1]
  (byte-varref *bar*)
  (byte-goto-if-nil TAG 8)
  (byte-varref *bar*)
  (byte-constant 2)
  (byte-plus)
  (byte-return)
  (TAG 8)
  (byte-constant foo)
  (byte-return)
\end{lstlisting}
where \verb|byte-varref| pushes the value of a symbol into the stack,
\verb|byte-goto-if-nil| pops the top of the stack and jumps to the
given label if it is \verb|nil|, \verb|byte-constant| pushes the value
from an immediate into the stack, \verb|byte-plus| pops two elements
from the stack, adds them up and pushes the result back into the stack
and \verb|byte-return| exits the function using the top of the stack
as return value.  An extensive description of these instructions is
available in Ref.~\cite[Sec.~1.3]{ELS:ELISP_bytecode}.

\textit{libgccjit} allows for describing code programmatically in
terms of \verb|gcc_jit_object|s created through C or C++
API~\cite[Sec.~\textit{Objects}]{ELS:libgccjit}.  The semantic it can
express can be described as a subset of the one of the C programming
language.  This includes \verb|l| and \verb|r| values, arithmetic
operators, assignment operators and function calls.  The most notable
difference with respect to C is that conditional statements such as
\verb|if| and \verb|else| are not supported and the code has to be
described in terms of basic blocks.  Inside GCC, \textit{libgccjit} IR
is mapped into GIMPLE when the actual compilation is requested.  One
key property of Emacs Lisp LAP is that it guarantees that, for any
given program counter, the stack depth is fixed and known at compile
time.  The previous LAP code can be transformed in the following
pseudo-code, suitable to be described in the \textit{libgccjit} IR:

\begin{lstlisting}[language=C, style=mystyle1]
  Lisp_Object local[2];

  bb_0:
    local[0] = varref (*bar*);
    if (local[0] == NIL) goto bb_2;
    else goto bb_1;

  bb_1:
    local[0] = varref (*bar*);
    local[1] = two;
    local[0] = plus (local[0], local[1]);
    return local[(int)0];

  bb_2:
    local[0] = foo;
    return local[0];
\end{lstlisting}

This transformation accomplishes the following:
\begin{itemize}
\item performs opcode decoding during the transformation so that it is
  not needed anymore at runtime.
\item decodes and compiles all the operations within the original
  stack into assignments.
\item splits the initial list of LAP instructions into basic blocks.
\end{itemize}
These tasks are performed by means of an intermediate translation into
LIMPLE, which enables standard code optimization routines present
in GCC as well as dedicated optimization passes.

\subsection{LIMPLE IR}

As previously introduced, in order to implement a series of
optimization passes, we defined an intermediate representation, that
we called ``LIMPLE'', whose main requirement is to be SSA.  The
description of every variable inside the compiler is accomplished with
instances of a structure we called \texttt{m-var} and reported in
Appendix~\ref{m-var}.  This represents the Lisp objects that will be
manipulated by the function being compiled.  A function in LIMPLE is a
collection of basic blocks connected by edges to compose a control
flow graph where every basic block is a list of {\verb|insn|}
(instructions).  The format of every LIMPLE {\verb|insn|} is a list
\verb|(operator operands)| whose valid operands depend on the
operator itself, such as:
\begin{itemize}
\item \texttt{(set dst src)} Copy the content of the slot represented
  by the \texttt{m-var} \texttt{src} into the slot represented by
  \texttt{m-var} \texttt{dst}.
\item \texttt{(setimm dst imm)} Similar to the previous one but
  \texttt{imm} is a Lisp object known at compile time.
\item \texttt{(jump bb)} Unconditional jump to basic block whose name
  is represented by the symbol \texttt{bb}.
\item \texttt{(cond-jump a b bb\_1 bb\_2)} Conditional jump to
  \texttt{bb\_1} if \texttt{a} and \texttt{b} are \texttt{eq} or to
  \texttt{bb\_2} otherwise.
\item \texttt{(call f a b \dots)} Call a primitive function \texttt{f}
  where \texttt{a,~b,~\dots} are \texttt{m-var}s used as parameters.
\item \texttt{(comment str)} Include annotation \texttt{str} as
  comment inside the \texttt{.eln} debug symbols (see Sec.~\ref{debug}).
\item \texttt{(return a)} Perform a function return having as return
  value the \texttt{m-var a}.
\item \texttt{(phi dst src1 \dots srcn)} Conventional $\Phi$ node used
  by SSA representation.  When all \texttt{m-var}s \texttt{src1 \dots
    srcn} have the same immediate value this is assigned to
  \texttt{dst}.  Similarly it happens for the type (no handling for
  type hierarchy is implemented up to date).
\end{itemize}

\section{Compilation passes}

The compilation pipeline is composed by a sequence of passes that,
starting from the input Elisp source code, apply a succession of
transformations to finally produce the executable code in the form of
a \verb|.eln| file.  The following sections describe each of the
compilation passes, all of them are implemented in Lisp with the
exception of \verb|final|.

However, before getting into the details of each pass, it is useful to
discuss the reason why the data-flow analysis and optimization
algorithms already present in the GCC infrastructure are not enough
for the Elisp semantic, and dedicated ones had to be developed in
Lisp.

\paragraph{Type propagation}

Emacs Lisp is a strong dynamically-typed programming language.  Typing
objects is done through tagging pointers~\cite{ELS:wiki_TagPoint}.
While GCC has passes to propagate both constants and ranges, it has no
visibility of the Lisp type returned by Lisp primitive functions and,
as a consequence, on the tag bits set.

\paragraph{Pure functions}

Similarly, GCC does not know which Lisp functions can be optimized at
compile time having visibility only on the local compilation unit.
Optimizable functions are typically pure functions or functions that
are pure for a specific set of parameters.

\paragraph{Reference propagation}

Another useful property to be propagated is if a certain object will
or will not be referenced~\cite{ELS:wiki_reference} during function
calls.  This information is required to generate a more efficient
code, as discussed in Sec.~\ref{sec:mvar_layout}.

\paragraph{Unboxing}

GCC does not offer infrastructure for unboxing values.  Although not
yet implemented, the proposed infrastructure is designed to host
further improvements, such as unboxing, requiring data-flow analysis
~\cite{ELS:wiki_boxing}.

\paragraph{Compiler hints}

The data-flow analysis can be fed with compiler hints about the type
of certain expressions, included as high-level annotations in the
source code by the programmer.

\paragraph{Warning and errors}

A data-flow analysis engine as the one proposed in this work could be
used in the future to provide more accurate warnings and errors during
the compilation phase.

\paragraph{GCC optimization constraints}

GCC optimization passes often adopt conservative strategies not to
break the semantic of all the supported programming languages.  As an
example, the GCC tail call optimization pass does not perform
transformations whenever an instruction referencing memory is present
in the compiled function.  Given the specific semantic of the code
generated by the proposed work, conditions as the one mentioned may be
too restrictive resulting in missed optimizations.

\subsection{\textit{spill-lap}}

As already discussed, the main input for the compilation process is
the Lisp Assembly Program Intermediate Representation (LAP IR).
\textit{spill-lap} runs the byte-compiler infrastructure with the
Elisp source as input collecting all top-level forms and spilling the
LAP before it is assembled into final byte-code.

\subsection{\textit{limplify}}
\label{sec:limpification}

This pass is responsible for translating LAP IR into LIMPLE IR. In
general, LAP is a sequence of instructions, labels and
jumps-to-label.  Since the Emacs byte-interpreter is a stack-based
machine, every LAP instruction manipulates the
stack~\cite{ELS:ELISP_bytecode},~\cite[Sec.~5.1]{ELS:Elisp_evo}.  It
is important to highlight that at this stage all the stack
manipulations performed by LAP instructions are compiled into a series
of {\verb|m-var|} assignments.  Spurious moves will eventually be
optimized out by GCC.  This pass is also responsible for decomposing
the function into lists of LIMPLE \verb|insn|s, or basic blocks.  The
code necessary for the translation of most LAP instructions is
automatically generated using the original instruction definition
specified in the byte-compiler.

\subsection{\textit{static single assignment (SSA)}}

This pass is responsible for bringing LIMPLE into minimal SSA form,
discussed in~\cite[Sec.~2.2]{ELS:SSA_book}, as follows:
\begin{enumerate}[a)]
\item Edges connecting the various basic blocks are created.
\item The dominator tree is computed for each basic
  block~\cite{ELS:Dominance}.
\item Dominator frontiers are computed for each basic block.
\item $\Phi$ functions are placed as described
  in~\cite[Sec.~3.1]{ELS:SSA_book}.
\item {\verb|m-var|}s goes through classic SSA renaming.
\end{enumerate}

Once LIMPLE is in SSA form every \verb|m-var| object appears as
destination of an instruction only in one place within the SSA
lattice.  The same object can be referenced multiple times as source
though, but each \verb|m-var| can be identified by its unique
\verb|id| slot.

\subsection{\textit{forward data-flow analysis}}

For each {\verb|m-var|}, this pass propagates the following properties
within the control flow graph: value, type and where the
{\verb|m-var|} will be allocated (see Sec.~\ref{sec:mvar_layout}).
Initially, all immediate values set at compile time by {\verb|setimm|}
are propagated to each destination {\verb|m-var|}.  Afterwards, for
each \verb|insn| in each basic block the following operations are
iteratively performed:
\begin{enumerate}[a)]
\item If the \verb|insn| is a $\Phi$, the properties of
  {\verb|m-var|}s present as source operands are propagated to the
  destination operand when in agreement.
\item If a function call has a known return type, this is propagated
  to the result.
\item If a function call to a pure function is performed with all
  arguments having a known value, the call is optimized out and the
  resulting value is substituted.
\item Assignments by {\verb|set|} operators are used to propagate all
  {\verb|m-var|}s.
\end{enumerate}

This sequence is repeated until no more changes are performed in the
control flow graph.

\subsection{\textit{call-optim}}
\label{sec:call_optim}
This pass is responsible for identifying all function calls to
primitives going through the {\verb|funcall|} trampoline and
substitute them with direct calls.
The primitive functions most commonly used in the original LAP
definition are assigned dedicated opcodes, as described in
~\cite[page~172]{ELS:ELISP_bytecode}.  When a call to one of these
functions is performed, the byte-interpreter can thus perform a direct
call to the primitive function.  All the remaining functions are
instead called through the {\verb|funcall|} trampoline, which carries
a considerable overhead.  This mechanism is due to the intrinsic limit
of the opcode encoding space.  On the other hand, native-compiled code
has the possibility to call all Emacs primitives without any encoding
space limitation.  After this pass has run, primitive functions have
all equal dignity, being all called directly irrespective of the fact
that they were originally assigned a dedicated byte-opcode or not.
The same transformation is performed for function calls within the
compilation unit when the compiler optimization level is set to its
maximum value (see Sec.~\ref{speed}).  This will improve the
effectiveness of inlining and other inter-procedural optimizations in
GCC.  Finally, recursive functions are also optimized to prevent
{\verb|funcall|} usage.

\subsection{\textit{dead-code}}

This pass cleans up unnecessary assignments within the function.  The
algorithm checks for all \texttt{m-var}s that are assigned but not
used elsewhere, removing the corresponding assignments.  This pass is
also responsible for removing function calls generated by compiler
type hints (see Sec.~\ref{type_hints}) if necessary.

\subsection{\textit{tail recursion elimination (TRE)}}

This peephole pass~\cite[Chap.~18]{ELS:Muchnick1997} performs a special
case of tail call optimization called tail recursion elimination.  The
pass scans all LIMPLE \verb|insn|s in the function searching for a
recursive call in tail position.  If this is encountered it is replaced
with the proper code to restart executing the current function using
the new arguments without activating a new function frame into the
execution stack.  This transformation is described
in~\cite[Chap.~15.1]{ELS:Muchnick1997}.

\subsection{\textit{final (code layout)}}

This pass is responsible for converting LIMPLE into \textit{libgccjit}
IR and invoking the compilation through GCC.  We point out that the
code we generate for native-compiled Lisp functions follows the same
ABI of Elisp primitive C functions.  Also, a minimal example of pseudo
C code for a native-compiled Elisp function is listed in
Appendix~\ref{sec:pseudo_c_example}.

\label{sec:mvar_layout}
When optimizations are not engaged, \texttt{m-var}s associated to each
function are arranged as a single array of Lisp objects.  This array
has the length of the original maximum byte-code stack depth.
Depending on the number of their arguments, Elisp primitive functions
present one of the following two C
signatures~\cite{ELS:Elisp_manual-primitives}:
\begin{enumerate}[a)]
\item \lstinline[basicstyle=\ttfamily\small]{Lisp_Obj fun (Lisp_Obj arg01, ..., Lisp_Obj argn)},\\
  for regular functions with a number of arguments known and smaller or equal to $8$.
\item \lstinline[basicstyle=\ttfamily\small]{Lisp_Obj fun (ptrdiff_t n, Lisp_Obj *args)},\\
  otherwise.
\end{enumerate}
where \verb|ptrdiff_t| is an integral type, \verb|n| is the number of
arguments and \verb|args| is a one-dimensional array containing their
values.  When a call of the second kind is performed, GCC clobbers all
the \verb|args| array content, regardless the number of arguments
\verb|n| involved in the call.  This means that the whole array
content after the call is considered potentially modified.  For this
reason the compiler cannot trust values already loaded in registers
and has to emit new load instructions for them.  To prevent this, when
the optimization we have called ``advanced frame layout'' is
triggered, each \texttt{m-var} involved in a call of the second kind
is rendered in a stack-allocated array dedicated to that specific
call.  All other \texttt{m-var}s are rendered as simple automatic
variables.  The advanced frame layout is enabled for every
compilation done with a non zero \verb|comp-speed|, as discussed in
Sec.~\ref{speed}.

This pass is also responsible for substituting the calls to selected
primitive functions with an equivalent implementation described in
\textit{libgccjit} IR.  This happens for small and frequently used
functions such as: \verb|car|, \verb|cdr|, \verb|setcar|,
\verb|setcdr|, \verb|1+|, \verb|1-|, or \verb|-| (negation).  As an
example, the signature for function \verb|car| implemented in
\textit{libgccjit} IR will be:

\begin{lstlisting}[language=C, style=mystyle1]
  static Lisp_Object CAR (Lisp_Object c,
                          bool cert_cons)
\end{lstlisting}

If compared to the original \verb|car| function a further parameter
has been added, \verb|cert_cons| which stands for ``certainly a
\verb|cons|''.  \textit{final} will emit a call to \verb|CAR| setting
\verb|cert_cons| to \verb|true| if the data-flow analysis was able to
prove \texttt{c} to be a \verb|cons| or setting it to \verb|false|
otherwise.  This mechanism is used in a similar fashion with most
inlinable functions injected by this pass in order to provide the
information obtained by the data-flow analysis to the GCC one.  Since
the GCC implementation has the full definition of these functions,
they can be optimized effectively.

\section{System Integration}

\subsection{Compilation unit and file format}

The source for a compilation unit can be a Lisp source file or a
single function and, as already mentioned, the result of the
compilation process for a compilation unit is a file with \verb|.eln|
extension.  Technically speaking, this is a shared library where Emacs
expects to find certain symbols to be used during load.  The
conventional load machinery is modified such that it can load
\verb|.eln| files in addition to conventional \verb|.elc| and
\verb|.el| files.

In order to be integrated in the existing infrastructure we define the
\verb|Lisp_Native_Comp_Unit| Lisp object.  This holds references to
all Lisp objects in use by the compilation unit plus a reference to
the original \verb|.eln|. Every \verb|.eln| file is expected to
contain a number of symbols including:
\begin{itemize}
\item \verb|freloc_link_table|: static pointer to a structure of
  function pointers used to call Emacs primitives from native-compiled
  code.
\item \verb|text_data_reloc|: function returning a string representing
  all immediate constants in use by the code of the compilation unit.
  The string is formed using \verb|prin1| so that it is suitable to
  be read by Lisp reader.
\item \verb|d_reloc|: static array containing the Lisp objects used by
  the compiled functions.
\item \verb|top_level_run|: function responsible of performing all
  the modifications to the environment expected by the load of the
  compilation unit.
\end{itemize}

\subsection{Load mechanism}

Load can be performed conventionally as: \verb|(load "test.eln")|.
Loading a new compilation unit translates into the following steps:

\begin{enumerate}[a)]
\item Load the shared library into the Emacs process address space.
\item Given that \verb|.eln| plugs directly into Emacs primitives,
  forward and backward version compatibility cannot be ensured.
  Because of that each \verb|.eln| is signed during compilation with
  an hash and this is checked during load.  In case the hash
  mismatches the load process is discarded.
\item Lookup the following symbols in the shared library and set their
  values: \verb|current_thread_reloc|, \verb|freloc_link_table|,
  \verb|pure_reloc|.
\item Lookup \verb|text_data_reloc| and call it to obtain the
  serialized string representation of all Lisp objects used by
  native-compiled functions.
\item Call the reader to deserialize the objects from this string and
  set the resulting objects in the \verb|d_reloc| array.
\item Lookup and call \verb|top_level_run| to have the environment
  modifications performed.
\end{enumerate}

We show in Appendix~\ref{sec:pseudo_c_example} an example of pseudo C
code for a native-compiled function illustrating the use of
\verb|freloc_link_table| and \verb|d_reloc| symbols.

When loaded, the native-compiled functions are registered as
\emph{subr}s as they share calling convention with primitive C
functions.  Both native-compiled and primitive functions satisfies
\verb|subrp| and are distinguishable using the predicate
\verb|subr-native-elisp-p|.

\subsection{Unload}

The unload of a compilation unit is done automatically when none of
the Lisp objects defined in it is referenced anymore.  This is
achieved by having the \verb|Lisp_Native_Comp_Unit| object been
integrated with the garbage collector infrastructure.

\subsection{Image dump}

Emacs supports dumping the Lisp image during its bootstrap.  This
technique is used in order to reduce the startup time.  Essentially a
number of \verb|.elc| files are loaded before dumping the Emacs image
that will be invoked during normal use.  As of Emacs~27 this is done
by default by relying on the portable dumper, which is in charge of
serializing all allocated objects into a file, together with the
information needed to revive them.  The final Emacs image is composed
by an executable plus the matching dump file.  Image dump capability
has been extended to support native-compiled code, the portable dumper
has been modified to be able to dump and reload
\verb|Lisp_Native_Comp_Unit| Lisp objects.

\subsection{Bootstrap}

Since the Elisp byte-compiler is itself written in Elisp, a bootstrap
phase is performed during the build of the standard Emacs
distribution.  Conventionally this relies on the Elisp
interpreter~\cite[Sec.~5.2.1]{ELS:Elisp_evo}.  We modified the Emacs
build system to allow for a full bootstrap based on the native
compiler.  The adopted strategy for this is to follow the conventional
steps to produce \verb|.eln| files instead of \texttt{.elc} when
possible (lexically scoped code) and fall-back to \verb|.elc| otherwise.
More than 700 Elisp files are native-compiled in this process.

\subsection{Documentation and source integration}

The function documentation string, \verb|describe-function|, and ``goto
definition'' mechanism support have been implemented and integrated
such that native-compiled code behaves as conventional byte-compiled
code.

\subsection{Verification}

A number of tests have been defined to check and verify the
compiler. These include some micro test cases taken from Tom Tromey's
JIT~\cite[Sec.~5.11]{ELS:Elisp_evo},~\cite{ELS:Tromey_blog}.  A
classical bootstrap compiler test has also been defined, where the
interpreted compiler is used to native-compile itself, and then the
resulting compiler is used to compile itself.  Finally, the two
produced binaries are compared. The test is successful if the two
objects are bytewise identical.

\section{Elisp Interface}

\subsection{Code optimization levels}
\label{speed}
Some special variables are introduced to control the compilation
process, most notably \verb|comp-speed|, which controls the
optimization level and safety of code generation as follows:
\begin{enumerate}
  \setcounter{enumi}{-1}
\item No optimization is performed.
\item No Lisp-specific optimization is performed.
\item All optimizations that do not modify the original Emacs Lisp
  semantic and safeness are performed. Type check elision is allowed
  where safe.
\item Code is compiled triggering all optimizations.  Intra
  compilation unit inlining and type check elision are
  allowed. User compiler hints are assumed to be correct and exploited
  by the compiler.
\end{enumerate}
\verb|comp-speed| also controls the optimization level performed by
the GCC infrastructure as indicated by the table below.

\begin{center}
 \begin{tabular}{l c c c c}
   \hline
   \verb|comp-speed| & 0 & 1 & 2 & 3 \\ [0.5ex]
   \hline
   propagate & n & n & y & y \\
   call-optim & n & n & y & y \\
   call-optim (intra CU) & n & n & n & y \\
   dead-code & n & n & y & y \\
   TRE & n & n & n & y \\
   advanced frame layout & n & y & y & y \\
   GCC -Ox & 0 & 1 & 2 & 3 \\
   \hline
\end{tabular}
\end{center}

\subsection{Language extensions}
\label{type_hints}

In order to allow the user to feed the data-flow analysis with
type suggestions, two entry points have been implemented:

\begin{itemize}
\item \verb|comp-hint-fixnum|
\item \verb|comp-hint-cons|
\end{itemize}

These can be used to specify that a certain expression evaluates to
the specified type. For example, {\verb|(comp-hint-cons x)|} ensures
that the result of the evaluation of the form itself is a \verb|cons|.
Currently, when \verb|comp-speed| is less or equal to 2, type hints
are compiled into assertions, while they are trusted for type
propagation when using \verb|comp-speed| 3. These low level primitives
are meant to be used to implement operators similar to Common Lisp
\verb|the| and \verb|declare|~\cite{ELS:ANSI_1992_DPA}.

\subsection{Debugging facility}
\label{debug}

\textit{libgccjit} allows for emitting debug symbols in the
generated code and dumping a pseudo C code representation of the
\textit{libgccjit} IR.  This is triggered for compilations performed
with \verb|comp-debug| set to a value greater than zero.  Debugging
the generated code is achieved using a conventional native debugger
such as \verb|gdb|~\cite{ELS:GDB_home}. In this condition, the
\verb|final| pass emits additional code annotations, which are visible
as comments in the dumped pseudo C code to ease the debugging (see
Appendix~\ref{sec:pseudo_c_example}).

\section{Performance improvement}

\begin{table*}
 \begin{tabular}{l c c c}
  \hline
  benchmark          & byte-comp runtime (s) & native-comp runtime (s) &   speed-up \\
  \hline
  inclist            &                 19.54 &                    2.12 &    $ 9.2x$ \\
  inclist-type-hints &                 19.71 &                    1.43 &    $13.8x$ \\
  listlen-tc         &                 18.51 &                    0.44 &    $42.1x$ \\
  bubble             &                 21.58 &                    4.03 &    $ 5.4x$ \\
  bubble-no-cons     &                 20.01 &                    5.02 &    $ 4.0x$ \\
  fibn               &                 20.04 &                    8.79 &    $ 2.3x$ \\
  fibn-rec           &                 20.34 &                    7.13 &    $ 2.9x$ \\
  fibn-tc            &                 21.22 &                    5.67 &    $ 3.7x$ \\
  dhrystone          &                 18.45 &                    7.22 &    $ 2.6x$ \\
  nbody              &                 19.79 &                    3.31 &    $ 6.0x$ \\
  \hline
  \\
 \end{tabular}
 \caption{Performance comparison of byte-compiled and native-compiled Elisp benchmarks}
 \label{perf_table}
 \vspace{-0.7cm}
\end{table*}

In order to evaluate the performance improvement of the native code, a
collection of Elisp benchmarks has been assembled and made available
as \verb|elisp-benchmarks| in the Emacs Lisp Package Archive
(ELPA)~\cite{ELS:bechmarks_package}.  It includes the following set of
programs:

\begin{itemize}
\item List processing: traverse a list incrementing all elements or
  computing the total length.
\item Fibonacci number generator: iterative, recursive and
  tail-recursive implementations.
\item Bubble sort: both destructive in-place and non destructive.
\item Dhrystone: the famous synthetic benchmark ported
  from C to Elisp~\cite{ELS:Weicker1984}.
\item N-body simulation: a model of the solar gravitation system,
  intensive in floating-point arithmetic.
\end{itemize}

The benchmarking infrastructure executes all programs in sequence,
each for a number of iterations selected to have it last around 20
seconds when byte-interpreted. The sequence is then repeated five
times and the execution times are averaged per each benchmark.  The
results reported in Table~\ref{perf_table} are obtained from an
Intel~i5--4200M machine.  They compare the execution time of the
benchmarks when byte-compiled and run under the vanilla Emacs~28 from
master branch against their native-compiled versions at
\verb|comp_speed| 3. The native-compiled benchmarks are run under
Emacs compiled and bootstrapped at \verb|comp_speed| 2 from the same
revision of the codebase.

The optimized native-code allows all the benchmarks to run at least
two times faster, with most of them reaching much higher performance
boosts. Despite the analysis being still preliminary, the reason
behind these improvements can be explained with several
considerations.  First of all the removal of the byte-interpreter loop
which, implementing a stack-machine, fetches opcodes from memory,
decodes and executes the corresponding operations and finally pushes
the results back to the memory.  Instead the native compiler walks the
stack at compile time generating a sequence of machine-level
instructions (the native code) that works directly with the program
data at execution time (see Sec.~\ref{sec:limpification}).  The result
of this process is that executing a native-compiled program takes a
fraction of machine instructions with respect to byte-interpreting
it. Analyzing the instructions mix also reveals a smaller percentage
of machine instructions spent doing memory accesses, in favor of data
processing ones.  This is the fundamental upgrade of native
compilation against interpretation and is probably the major source of
improvement for benchmarks with smaller speed-ups, where no other
optimizations apply.

On the other hand, benchmarks with larger improvements also take
advantage of Lisp specific compiler optimizations, in particular from
call optimizations (see Sec.~\ref{sec:call_optim}).  Function calls
avoid the trampoline when targeting subroutines defined in the same
compilation unit or when calling pure functions from the C codebase.
Moreover, the data-flow analysis step allows to exploit the properties
of the structures manipulated by the compiler in order to produce code
with less overheads.  Function calls can also be completely removed,
along with corresponding returns, and replaced by simple jumps for
optimized tail recursive functions, avoiding at the same time new
allocations on the execution stack.

Finally, the data-flow analysis can be made even more effective when
paired with compiler hints.  Without these, the only types known at
compile time are the ones belonging to constants or values returned by
some primitive functions.  Type hints greatly increase the chances for
the native compiler to optimize out expensive type checks.  In our
measurements, the same benchmark (\textit{inclist}) annotated with
type hints earns a further improvement of $50\%$ in terms of execution
speed, while compiled under the same conditions.

\section{Conclusions}

In this work we discussed a possible approach to improve execution
speed of generic Elisp code, starting from LAP representation and
generating native code taking advantage of the optimization
infrastructure of the GNU Compiler Collection.  Despite its early
development stage, the compiler successfully bootstraps a usable Emacs
and is able to compile all lexically scoped Elisp present in the Emacs
distribution and in ELPA.  The promising results concerning stability
and compatibility already led this work to be accepted as
\textit{feature~branch} in the official GNU Emacs repository.
Moreover, a set of benchmarks was developed to evaluate the
performance gain and preliminary results indicate an improvement of
execution speed between 2.3x and 42x, measured over several runs.  At
last, we point out that most of the optimization possibilities allowed
by this infrastructure are still unexplored. Already planned
improvements include: supporting \verb|fixnum| unboxing and full tail
call optimization, exposing more primitives to the GCC infrastructure
by describing them in the \textit{libgccjit} IR during the
\verb|final| pass, and allowing to signal warnings and error messages
at compile time based on values and types inferred by the data-flow
analysis.

\begin{acks}
  Stefan Monnier and Kyrylo Tkachov for reviewing this manuscript.
  Luca Benso for inspiration, useful discussions and artichokes, and
  Daniela Ferraro for cooking them.
\end{acks}

\bibliographystyle{plainnat}
\bibliography{references}

\onecolumn
\appendix

\section{Definition of m-var}
\label{m-var}

\begin{lstlisting}[language=lisp, style=mystyle2]
(cl-defstruct (comp-mvar (:constructor make--comp-mvar))
    "A meta-variable being a slot in the virtual-stack."
    (id nil :type (or null number)
        :documentation "SSA unique id number when in SSA form.")
    (const-vld nil :type boolean
        :documentation "Validity signal for the following slot.")
    (constant nil
        :documentation "When const-vld is non-nil this is used for holding a known value.")
    (type nil
        :documentation "When non-nil indicates the type known at compile time."))
\end{lstlisting}

\vspace{.3cm}

\section{Example of a compilation unit}
\label{sec:pseudo_c_example}

Below we show an example of an elementary compilation unit followed by
the pseudo C code generated dumping the \textit{libgccjit} IR.  The
compilation process is performed using \texttt{comp-speed = 3} and
\texttt{comp-debug = 1}.

\vspace{.3cm}

\begin{lstlisting}[language=lisp, style=mystyle1]
;;;  -*- lexical-binding: t -*-
(defun foo ()
  (if *bar*
      (+ *bar* 2)
      'foo))
\end{lstlisting}

\vspace{.3cm}

\begin{lstlisting}[language=C, style=mystyle1]
extern union comp_Lisp_Object
F666f6f_foo ()
{
  union comp_Lisp_Object[2] arr_1;
  union comp_Lisp_Object local0;
  union cast_union union_cast_28;
entry:
  /* Lisp function: foo */
  goto bb_0;
bb_0:
  /* const lisp obj: *bar* */
  /* calling subr: symbol-value */
  local0 = freloc_link_table->R73796d626f6c2d76616c7565_symbol_value (d_reloc[0]);
  /* const lisp obj: nil */
  union_cast_28.v_p = (void *)NULL;
  /* EQ */
  if (local0.num == union_cast_28.lisp_obj.num) goto bb_2; else goto bb_1;
bb_2:
  /* foo */
  local0 = d_reloc[2];
  /* const lisp obj: foo */
  return d_reloc[2];
bb_1:
  /* const lisp obj: *bar* */
  /* calling subr: symbol-value */
  arr_1[0] = freloc_link_table->R73796d626f6c2d76616c7565_symbol_value (d_reloc[0]);
  /* const lisp obj: 2 */
  arr_1[1] = d_reloc[3];
  /* calling subr: + */
  local0 = freloc_link_table->R2b_ (2, (&arr_1[0]));
  return local0;
}
\end{lstlisting}

\end{document}